\newif\ifmnras
\mnrastrue
\ifmnras
	\documentclass[fleqn,usenatbib]{mnras}
\else
	\documentclass[apj,numberedappendix]{emulateapj}
\fi

\usepackage{graphics,epsf}
\usepackage{amsmath}                
\usepackage{amsfonts}               
\usepackage{amssymb}                
\usepackage{epsfig}                 
\usepackage{graphicx}

\def \cm{~\rm{cm}}
\def \s{~\rm{s}}
\def \km{~\rm{km}}

\def \K{~\rm{K}}
\def \g{~\rm{g}}

\def \AU{~\rm{AU}}
\def \erg{~\rm{erg}}

\def \yr{~\rm{yr}}

\def \dday{~\rm{day}}

\ifmnras
	\def \aap{A\&A}
	
	\def \apj{ApJ}
	\def \apjl{ApJ}
	\def \apjs{ApJS}

	\def \mnras{MNRAS}

\fi

\usepackage{xcolor}
\definecolor{redak}{rgb}{0.9,0.15,0.05}

\ifmnras

\title[Energy transport by convection in the CEE]
{Energy transport by convection in the common envelope evolution}

\author	[Sabach, Hillel, Schreier, Soker]{
Efrat Sabach, Shlomi Hillel, Ron Schreier, and Noam Soker\thanks{E-mail: efrats@physics.technion.ac.il, shlomihi@tx.technion.ac.il, ronsr@physics.technion.ac.il, soker@physics.technion.ac.il}
\\
Department of Physics, Technion -- Israel
Institute of Technology, Haifa 32000 Israel\\
}

	\date{Accepted XXX. Received YYY; in original form ZZZ}

	\pubyear{2017}
\fi

\begin{document}
\label{firstpage}

\ifmnras
	\pagerange{\pageref{firstpage}--\pageref{lastpage}}
	\maketitle
\else
	\title{?}

	\author{Efrat Sabach}
	\author{Noam Soker}
	\affil{Department of Physics, Technion -- Israel
	Institute of Technology, Haifa 32000, Israel;
	efrats@physics.technion.ac.il; soker@physics.technion.ac.il}
\fi

\begin{abstract}
We argue that outward transport of energy by convection and photon diffusion in a common envelope evolution (CEE) of giant stars substantially reduces the fraction of the recombination energy of hydrogen and helium that is available for envelope removal.
We base our estimate on the properties of an unperturbed asymptotic giant branch (AGB) spherical model, and on some simple arguments. Since during the CEE the envelope expands and energy removal by photon diffusion becomes more efficient, our arguments underestimate the escape of recombination energy. We hence strengthen earlier claims that recombination energy does not contribute much to common envelope removal. 
A large fraction of the energy that jets deposit to the envelope, on the other hand, might be in the form of kinetic energy of the expanding and buoyantly rising hot bubbles. These rapidly rising bubbles remove mass from the envelope. We demonstrate this process by conducting a three-dimensional hydrodynamical simulation where we deposit hot gas in the location of a secondary star that orbits inside the envelope of a giant star.
Despite the fact that we do not include the large amount of gravitational energy that is released by the in-spiraling secondary star, the hot bubbles alone remove mass at a rate of about $0.1 M_\odot \yr^{-1}$, which is much above the regular mass loss rate.   
\end{abstract}



\section{INTRODUCTION}
\label{sec:intro}
Even with more and more sophisticated three-dimensional simulations of
the common envelope evolution (CEE) the community did not yet produce
the expected results of full envelope ejection (e.g., \citealt{LivioSoker1988,
RasioLivio1996, SandquistTaam1998, Sandquistetal2000, Lombardi2006, RickerTaam2008, TaamRicker2010, DeMarcoetal2011, Passyetal2011, Passyetal2012, RickerTaam2012, Nandezetal2014,
Ohlmannetal2016, Ohlmannetal2016b, Staffetal2016MN8, NandezIvanova2016,
Kuruwitaetal2016, IvanovaNandez2016, Iaconietal2017, DeMarcoIzzard2017, Galavizetal2017, Iaconietal2017a}).
These difficulties arise although in most cases the gravitational energy that is released by the in-spiraling binary system is larger than the binding energy of the common envelope (CE; e.g., \citealt{DeMarcoetal2011, NordhausSpiegel2013}). It seems that this is not a numerical problem, but rather there is a fundamental problem for the binary system to release the orbital energy when the orbital separation is small. The reason might be that there is only little envelope mass at small orbital separations \citep{Soker2013}. 

To overcome the problem of envelope removal, it has been suggested that either the rotating envelope has an enhanced mass loss rate due to radiation pressure on dust \citep{Soker2004, Soker2017final}, or that extra energy sources exist.
One such extra energy source might be jets that are launched by the secondary star, being a main sequence (MS) star or a more compact object (e.g., \citealt{Soker2014, Shiberetal2016, MorenoMendezetal2017}).
This is supported in part by the finding of \cite{BlackmanLucchini2014} who deduce from the momenta of bipolar PNe that strongly interacting binary systems, probably in a CEE, can launch energetic jets. \cite{ArmitageLivio2000} and \cite{Chevalier2012} already studied the ejection of the CE by jets launched from a neutron star (NS) companion, but they did not extend the jet-mechanism to include other types of secondary stars.

Another extra energy source to remove the CE that has been proposed is the recombination energy of hydrogen and helium (e.g., \citealt{Nandezetal2015, NandezIvanova2016} for recent papers, and \cite{Kruckowetal2016} for a recent discussion of recombination and accretion extra energy sources ).
For efficient removal of the CE by recombination energy, it is required that there will be no time for radiation to escape and carry this energy out of the envelope. \cite{Harpaz1998} and \cite{SokerHarpaz2003} argued that the sharp reduction of the optical depth in the hydrogen recombination zone allows the radiation to escape, and hence the contribution of hydrogen recombination energy to the CE ejection is very small.
For example, if the recombination takes place at an optical depth of $\tau \approx 100-1000$
(e.g., \citealt{Ivanovaetal2015, NandezIvanova2016}), the photon diffusion time is much shorter
than the acceleration time of the wind, and most photons diffuse out. Some recent suggestions that the recombination energy is an important energy source (\citealt{NandezIvanova2016, IvanovaNandez2016}) to the removal of the CE do not confront these arguments.

In the present study we examine the time it requires for radiation and convection to carry recombination energy out of an asymptotic giant branch (AGB) star, and from that put an upper limit on the efficiency by which the recombination energy can be used to expel the CE.
Giant stars, such as red giant branch (RGB) and AGB stars, have very strong convection. In general, however, three dimensional (3D) numerical simulations of the CEE do not include the energy that can be carried out by convection. The reason for this is numerical limitations.
Earlier calculations of the energy transport by convection during the CEE
(e.g., \citealt{MeyerHofmeister1979, Podsiadlowski2001, Ivanovaetal2015}) 
concentrated on the later CEE phases, and did not consider the removal of recombination energy. We here concentrate on removal of recombination energy. 

\section{RADIATIVE COOLING OF RECOMBINATION LAYERS}
\label{sec:radiative}

To study the diffusion in a typical AGB star we conduct stellar evolution simulations using the \texttt{MESA} (Modules for Experiments in Stellar Astrophysics) code,
version 9575 (\citealt{Paxtonetal2011, Paxtonetal2013, Paxtonetal2015}).
We calculate stellar evolution from zero age main sequence (ZAMS) until the formation of a white dwarf (WD) for a star with a ZAMS mass of 
$M_{1,\rm ZAMS}=2 M_{\sun}$ and with solar metallicity, $Z=0.02$.

The photon diffusion time out of a recombination zone at a depth of $\Delta R$ below the photosphere is 
\begin{equation}
t_{\rm diff} \approx \frac{ 3 \tau \Delta R}{c} =
1.1 
 \left( \frac{\tau}{2.5\times 10^5} \right)
  \left( \frac{\Delta R}{20 R_\odot} \right)
\label{eq:tdiff} \yr,
\end{equation}
where $\tau$ is the optical depth from the recombination zone outwards, and $c$ is the speed of light.
We scaled the values of $\tau$ and $\Delta R$ with the typical quantities of the zone where the ionization fraction of hydrogen is $50 \%$ in our AGB model of mass of $M_{1,\rm AGB}=1.75 M_\odot$, and a radius of $R_{1}=220 R_\odot$. 
We present the optical depth and diffusion time for this model in the upper panel of Fig. \ref{fig:t_diff,tau_Mi=2}. 
In the lower panel we present the ionization degrees of hydrogen and helium. 
\begin{figure}
\begin{center}
\vspace*{-1cm}
\hspace*{-0.5cm}
\includegraphics[width=9.2cm]{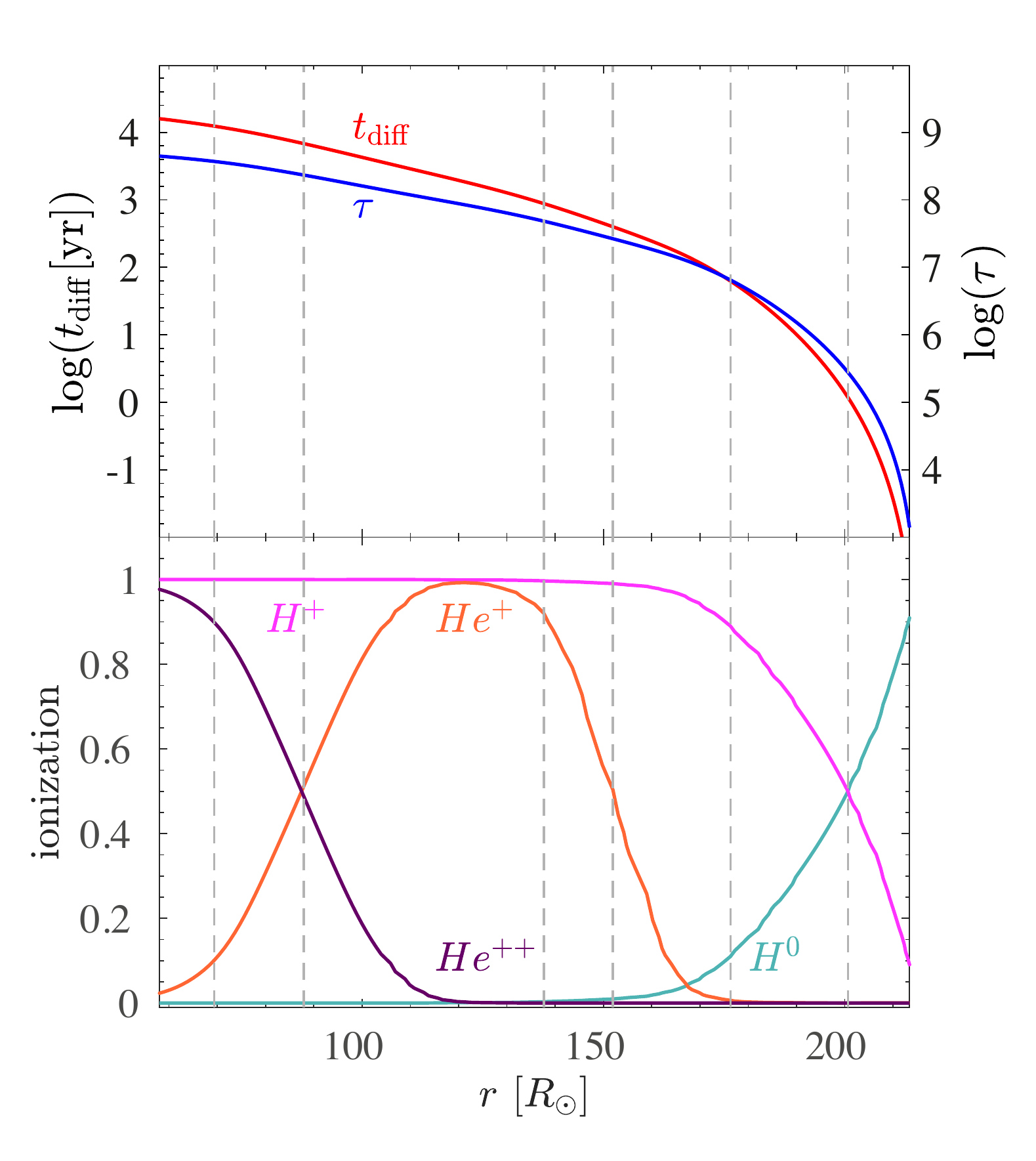}
\vspace*{-0.8cm}
\caption{
Some profiles in the outer envelope of an AGB star of mass $M_{1}=1.75 M_\odot$ and a radius of $R_{1}=220R_\odot$.
Its zero age main sequence mass was $M_{\rm 1,ZAMS}=2M_\odot$.
Upper panel: The diffusion time $t_{\rm diff}$ (red) and the optical depth $\tau$ (blue).
Lower panel: The ionization degree of H$^0$, H$^+$, He$^+$, and He$^{++}$.
The gray lines are to guide the eye to the different ionization levels;  from left to right: 
He$^{++}$ at 90\%, He$^{++}$ at 50\%, He$^{+}$ at 90\%, He$^{+}$ at 50\%, H$^{+}$ at 90\%, and
H$^{0}$ at 50\%.}
\label{fig:t_diff,tau_Mi=2}
\end{center}
\end{figure}

The time scale to deposit energy into the envelope during its removal is
the plunge-in time of the secondary star plus the acceleration time of the envelope.
According to the results of, e.g., \cite{Passyetal2012} and \cite{Ohlmannetal2016}, the plunge-in time is a little below the orbital time on the surface of the giant $T_{\rm orb}$. We therefore take the envelope ejection time to be 
$t_{\rm ej} \approx T_{\rm orb}={2\pi} R_1 /v_{\rm kep}$, where $v_{\rm kep}$ is the Keplerian velocity on the surface of the giant at radius $R_1$.

The ratio $F_\gamma$ of photon energy that is used to accelerate the gas is crudely given by 
\begin{eqnarray}
\begin{aligned}
F_\gamma < & \left( 1+ \frac{t_{\rm ej}}{t_{\rm diff}} \right)^{-1} 
 \\
= &  \left[ 1+  0.6 
 \left( \frac{\Delta R}{0.1 R_1} \right)^{-1}
 \left( \frac{\tau}{2.5\times 10^5} \right)^{-1}
 \left(  \frac{v_{\rm Kep}(R_1)}{40 \km \s^{-1}} \right)^{-1} \right]^{-1},
\label{eq:Fgamma}
\end{aligned}
\end{eqnarray}
where we scale quantities according to the zone where the ionization degree of hydrogen is $50 \%$. 
As the star expands, the diffusion time scale decreases and the expansion time increases. Therefore, the fraction $F_\gamma$ given above is an upper limit. 

The value of $F_\gamma ({\rm H}^+)$ implies that a substantial fraction of the energy carried by hydrogen recombination photons ends in radiation rather than in mechanical energy of the envelope \citep{Harpaz1998}. 

As evident from the high optical depth of the helium ionizations zones, the radiation by itself has no time to remove the energy from helium recombination if that zone expands on a dynamical time. 

\section{CONVECTIVE COOLING OF RECOMBINATION LAYERS}
\label{sec:convective}

We now examine whether convection might carry the recombination energy to the photosphere, where it will radiate away. We present the relevant quantities, density and sound speed in the envelope, in the upper panel of Fig.  \ref{fig:rho,c_s_Mi=2} for our AGB model (see section \ref{sec:radiative}). In the lower panel we present again the ionization degrees of hydrogen and helium. 
\begin{figure}
\begin{center}
\vspace*{-0.2cm}
\hspace*{-0.5cm}
\includegraphics[width=9.2cm]{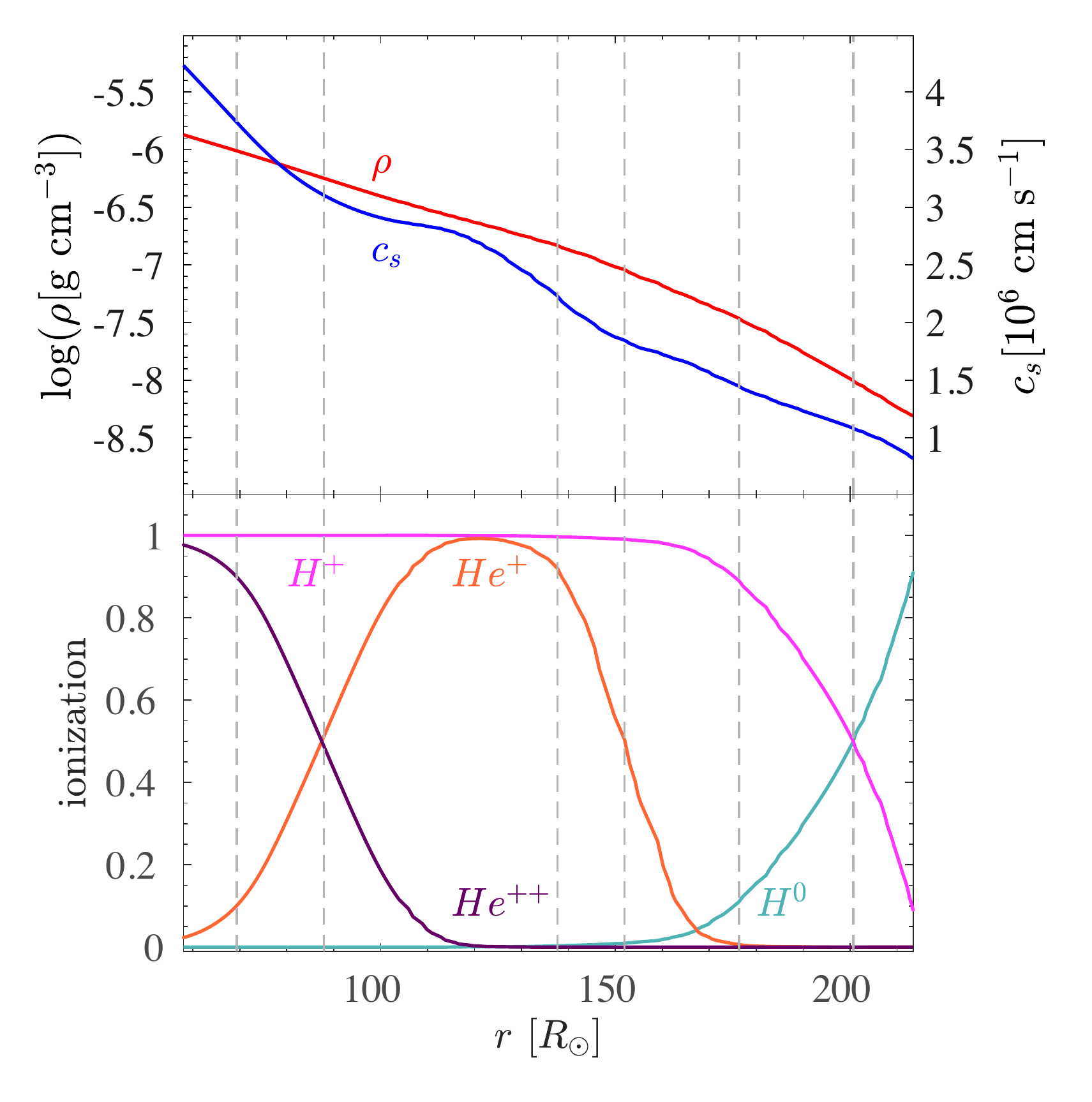}
\vspace*{-0.8cm}
\caption{
Upper panel: The profile of the density $\rho$ (red) and the sound speed $C_{\rm s}$ (blue) in the convective envelope of an AGB star with a zero age main sequence mass of $M_{\rm 1,ZAMS}=2M_\odot$. 
It is the same model as presented in Fig. \ref{fig:t_diff,tau_Mi=2}, having a mass and radius of $M_1=1.75 M_\odot$ and $R_1=220R_\odot$, respectively.
Lower panel is as in Fig. \ref{fig:t_diff,tau_Mi=2}.
}
\label{fig:rho,c_s_Mi=2}
\end{center}
\end{figure}

\cite{Ohlmannetal2016} conducted 3D simulations of the CEE, and found instabilities that indicate the onset of turbulent convection.
They further noted that these can play a significant role in the energy transport on thermal time scales.
We here present a crude estimate and show that energy transport by convection cannot be neglected when the helium recombines.

Consider an envelope with a density profile of $\rho(r) = \rho_{\rm rec} (r/r_{\rm rec})^{-2}$, 
which is rather a reasonable approximation for large AGB stars, i.e., radii of $R_\ast \ga 1\AU$, in the helium ionization zone.
Let $r_{\rm rec}$ be the radius where the recombination takes place, i.e., ionization fraction is about 50 per cent. 
The envelope mass inner to this radius is $M_{\rm env}(r_{\rm rec}) = 4 \pi \rho_{\rm rec} r^3_{\rm rec}$.
The energy that can be released by recombination of He$^{++}$ to He$^{+}$ of that mass is
\begin{equation}
\begin{aligned}
E & ({\rm He}^{++}, r_{\rm rec})=   Y \frac{M_{\rm env}(r_{\rm rec})}{m_{\rm He}} 54.4 {~\rm ev}  \\
            =  & 6.1 \times 10^{45} 
\left( \frac{\rho_{\rm rec}}{5 \times 10^{-7} \g \cm^{-3}} \right)
\left( \frac{r_{\rm rec}}{90 R_\odot} \right)^3 \erg.
\label{eq:Erec++}
\end{aligned}
\end{equation}
For these parameters the envelope mass inward to radius $r$ is $M_{\rm env}(r) = 0.776(r/r_{\rm rec}) M_\odot$. 
We take a helium mass fraction of $Y=0.3$ and we scale quantities with their typical values in the zone where the ionization fraction of He$^{++}$ is 50 per cent.

For the maximum power that subsonic convection can carry we use the expression
given by \cite{QuataertShiode2012} and \cite{ShiodeQuataert2014}
\begin{equation}
\begin{aligned}
L_{\rm max,conv}(r_{\rm rec}) = & 4 \pi \rho(r) r^2 c_s^3(r)                       = 1.7 \times 10^{6}  
\left( \frac{\rho_{\rm rec}}{5 \times 10^{-7} \g \cm^{-3}} \right) \\ 
 \times & 
 \left( \frac{r_{\rm rec}}{90 R_\odot} \right)^2
\left( \frac{c_s}{30 \km \s^{-1}} \right)^3 L_\odot,
\label{eq:lmaxconv1}
\end{aligned}
\end{equation}
If the entire He$^{++}$ in the mass $M_{\rm env}(r_{\rm rec})$ recombines, the convection can carry this energy out of the recombination region in a time of
\begin{equation}
\tau_{\rm min, conv} (r_{\rm rec})= \frac {E ({\rm He}^{++}, r_{\rm rec})}{L_{\rm max,conv}(r_{\rm rec})} =11
\left( \frac{r_{\rm rec}}{90 R_\odot} \right)
\left( \frac{c_s}{30 \km \s^{-1}} \right)^{-3} \dday.
\label{eq:THe++}
\end{equation}

The time it requires for the convection to carry this energy out to 
an optically thin region where it can be radiated away is much longer. 
The convective cells need to move a distance  $\Delta R$ 
from the recombination zone to the photosphere. 
They do it in an average sound speed $\bar c_s$.
The energy transport time is then
\begin{equation}
\tau_{\rm out, conv} \simeq  50
\left[ \frac{\Delta R({\rm He}^{++})}{130 R_\odot} \right]
\left( \frac{\bar c_s}{20 \km \s^{-1}} \right)^{-1} \dday,
\label{eq:Tout}
\end{equation}
The relevant time scale of recombination in the simulation of \cite{IvanovaNandez2016} is about 100-200 days. We therefore conclude that a large fraction, possibly more than half, of the recombination energy of the He$^{++}$ can be carried away by convection, and then radiated away.

The recombination of He$^{+}$ to neutral helium and of H$^+$ to hydrogen in their simulations lasts for a longer time than the recombination time of He$^{++}$ to He$^{+}$. In addition, the flow time of convective cells from the recombination zones of He$^+$ and H$^+$ to the photosphere is shorter than the convection motion from the recombination zone of He$^{++}$ to the photosphere.
It turns out that the fraction of the recombination energy of He$^+$ and H$^+$ that is radiated away is larger than that of the He$^{++}$ recombination energy.
Overall, less than a half of the total recombination energy of helium, and much less than half of the recombination energy of hydrogen, are available to eject the envelope.

We calculate the binding energy of the envelope for the density profile
$\rho(r) = \rho_{\rm rec} (r/r_{\rm rec})^{-2}$, as used in deriving equation \ref{eq:Erec++}.
Including the virial theorem, the binding energy of the envelope is taken to be about half the magnitude of its gravitational energy. Integration over spherical shells from an inner envelope radius of $R_{\rm env,in}$, which here we take to be about $1 R_\odot$, to radius $r$ gives the approximate expression for the binding energy of the envelope inner to radius $r$ 
\begin{equation}
\begin{aligned}
E_{\rm bin} (r) & \simeq  2 \pi G \rho_{\rm rec} r_{\rm rec}^2 \left[ M_{\rm env}(r_{\rm rec})\frac{r}{r_{\rm rec}} + M_{\rm core}
    \ln \left( \frac{r}{R_{\rm env,in}} \right) \right]  
   \\  = & 5 \times 10^{46} 
    \left( \frac{\rho_{\rm rec}}{5 \times 10^{-7} \g \cm^{-3}} \right)
\left( \frac{r_{\rm rec}}{90 R_\odot} \right)^2 
   \frac{1}{3.18}  \\
 \times & \left[ 0.78 \frac{M_{\rm env}(r_{\rm rec})}{0.78 M_\odot}\frac{r}{r_{\rm rec}} + 
    2.4 \frac{M_{\rm core}}{0.6 M_\odot}
    \frac{ \ln (\frac{r}{R_{\rm env,in}})}{4}  \right] \erg, 
\label{eq:Ebin}
\end{aligned}
\end{equation}
where $M_{\rm env}(r_{\rm rec})$ is the envelope mass inner to $r_{\rm rec}$; for the parameters used here $M_{\rm env}(r_{\rm rec})=0.776 M_\odot$. 

The model we simulated using \texttt{MESA} as described in section \ref{sec:radiative} and presented in the figures there, has a core radius of $r_{\rm core}=0.03 R_\odot$, a radius of 50 per cent He$^{++}$ ionization fraction of $r_{\rm rec} = 88 R_\odot$, an envelope mass inward to this radius of $M_{\rm env}(r_{\rm rec})= 0.63 M_\odot$, and a core mass of $M_{\rm core} = 0.55 M_\odot$. We find the binding energy of the mass inward to radius $r_{\rm rec}$ and down to $R_{\rm env,in}=10\cdot r_{\rm core} =0.3 R_\odot$  to be $E_{\rm bin} = 2 \times 10^{46} \erg$. 

Overall, we find that the total energy that can be liberate by the recombination of He$^{++}$ to He$^{+}$ (in the entire region inward to $r_{\rm rec}$) is a small fraction of the envelope binding energy, $E ({\rm He}^{++}, r_{\rm rec}) \simeq 0.25 E_{\rm bin} (r_{\rm rec})$. When we consider the removal of abut half of this energy by convection and then radiation, and that the envelope leaves with positive energy (and not zero energy), we conclude that the recombination of He$^{++}$ might at most contribute about 10 per cent of the envelope removal energy. 

We conclude that most of the recombination energy is radiated away. The radiation is added  to the luminosity of the event, which might be classified as an intermediate luminosity optical transient (ILOT) event, that is expected at the beginning of the CEE (e.g. \citealt{RetterMarom2003, Retteretal2006, Tylendaetal2011, Tylendaetal2013, Nandezetal2014, Zhuetal2016, Soker2016GEE, Galavizetal2017}). 

\section{ENERGY DEPOSITION BY JETS}
\label{sec:jets}
\subsection {Preface}
\label{subsec:preface}
Based on section \ref{sec:convective}, we argue that in many cases the energy carried by convection during the CEE reduces the efficiency by which the energy released by recombination and the spiraling-in process can be used to eject the envelope.
The entropy of the rising convective cells in the envelope is not much higher than the entropy of their surrounding, and eventually they reach a radius where they deposit their energy and mix with the surrounding envelope.

When considering a common envelope of a giant star with a main sequence companion, the situation is likely to be different. As the secondary accretes mass it launches two opposite jets.   When the gas in the jets hits the envelope and passes through a shock wave it inflates bubbles. Jets are launched at velocities of about the escape speed from the secondary star, whether a MS star, a WD or a NS. For MS stars the jet velocity is then $v_j \simeq 500 \km \s^{-1}$. As long as the secondary star is not too close to the core of the giant star,  e.g., it is at an orbital separation of $a \ga 5 R_\odot$, the jets' velocity is much higher than the relative velocity of the secondary star and the envelope, and much larger than the velocity of the convective cells.
When the gas in jets is shocked it inflates high-temperature bubbles, with a typical temperature of $T_{\rm bub} \approx 3 \times 10^6 \K$.
The entropy of the inflated bubbles is much higher than that of the envelope. 

The hot bubbles are expected to rise buoyantly and accelerate envelope gas outwards.
The expanding jets and the rising bubbles carry mostly kinetic energy. In the recombination process, on the other hand, most of the energy is thermal, and convection can carry a large fraction of it outwards (section \ref{sec:convective}).

To study the processes described above, we perform a 3D simulation of a MS star that orbits inside the envelope of a giant star and launches jets. 

\subsection {The numerical scheme}
\label{subsec:scheme}

We run the stellar evolution code \texttt{MESA} to obtain a spherical AGB model with ZAMS mass of $M_{1, \rm ZAMS}=4 M_{\sun}$. We let the star evolve until it reaches the AGB stage after $3\times10^8 \yr$. At that time the stellar mass is $M_{1}=4 M_\odot$, its radius is $R_{1}=100\,R_{\sun}$, and its effective temperature is $T_{1, \rm eff}= 3400 \K$.
This stellar model is the same one we have used for our simulations of instantaneously energy injection simulations \citep{Hilleletal2017}.  
We then import the spherical AGB model, namely, the profiles of density and pressure, into the three-dimensional (3D) hydrodynamical code {\sc pluto} \citep{Mignone2007}. 
The full 3D Cartesian grid is taken as a cube with side lengths of $400\,R_{\sun}$. The center of the AGB star is placed at the center of the grid at $(x,y,z)=(0,0,0)$.
We employ an adaptive mesh refinement (AMR) grid with four refinement levels.
We use an equation of state of an ideal gas with adiabatic index $\gamma=5/3$. 
The base grid resolution is $1/48$ of the grid length (i.e., $8.33 R_{\sun}$), and the highest resolution is $2^3$ times smaller (i.e., $\sim 1 R_{\sun}$). The refinement criterion is the default AMR criterion in {\sc pluto} v. 4.2, based on the second derivative error norm \citep{Lohner1987} of the total energy density, which effectively tracks the secondary star and the perturbed regions.
In one test we have reduced the number of AMR refinements levels from 4 to 3. The results of the simulation were similar to those of the productive run. 
At this point our computer resources do not allow us to increase the AMR refinements levels from 4 to 5.

We do not simulate the spiraling-in of the secondary star,
but rather let the secondary star orbit at a constant orbital separation of $a=50\, R_{\sun}$ inside the envelope, where the density is $\rho_{50} = 1.06 \times 10^{-5} \g \cm^{-3}$. 
The mass of the AGB star inwards to $r=50\, R_{\sun}$ is $M_{1,50}=2.7 M_{\sun}$, the Keplerian velocity at that radius for a very low mass secondary star is $v_{\rm Kep,50} = 102 \km \s^{-1}$ and the orbital period is about 25 days. 
The Bondi-Hole-Lyttleton (BHL) accretion radius for these parameters is $R_{BHL} \simeq 2GM_2/v^2 \simeq 10 R_{\sun}$, where here $v=(v^2_{\rm rel} + c^2_s)^{1/2}$, $v_{\rm rel}$ is the relative velocity of the secondary star and the envelope, $c_s$ is the sound speed in the envelope, and we substituted $M_2=0.3 M_{\sun}$. Since the envelope is expected to rotate, $v_{\rm rel}<v_{\rm Kep}$. 
The BHL mass accretion rate for $M_2=0.3 M_\odot$ and our AGB model at $a=50 R_{\sun}$ is 
$\dot{M}_{\rm BHL} \simeq \pi R_{BHL}^2 v \rho = 2.7\,M_{\sun}/yr$.

As numerical calculations show that the accretion rate in the envelope is lower than the BHL value (e.g. \citealt{RickerTaam2012, MacLeodRamirezRuiz2015}), and disk formation might be inefficient (e.g., \citealt{MurguiaBerthieretal2017}), we take the actual accretion rate to be only 
$\dot M_{\rm acc}= 0.05 \dot M_{\rm BHL}=0.135 M_{\sun} \yr^{-1}$. 
We assume that the mass outflow rate in the two jets is 
$\dot M_{2j}=0.1 \dot M_{\rm acc} = 0.005 \dot M_{\rm BHL}$.
The power carried by the jets in our simulation is therefore 
$\dot{E_{2j}} = 0.005 \dot{M}_{\rm BHL}\cdot v^2_j/2
=1 \times 10^{39} \erg \s^{-1}$. 

This power is super-Eddington. However, we note that the flow into and out of the secondary star is not spherical. Mass inflows from the equatorial plane direction, and mass and energy are lost along the polar directions. Actually, the jets are likely to carry all the extra energy released by the accreted mass. As the power for the accretion rate found above is $\dot E_{\rm acc} \simeq 8 \times 10^{39} \erg \s^{-1}$, we could do with an accretion rate as low as about $0.01 \dot M_{\rm BHL}$. 

Due to limitations in the numerical resolution we cannot launch jets and follow the formation of hot bubbles by the shocked jets' gas (for the dynamics of jets in the CEE see \citealt{MorenoMendezetal2017}).
The reason for this limitation is that when the outflow from the secondary star is not spherical, we need to take care of the equatorial region around the secondary star where there is no outflow. This requires some more resolution, hence it demands more computing resources, in particular for the simulated cases when the secondary moves through the envelope (i.e., the ambient gas is not static).
Instead, we insert isotropic wind that is immediately shocked and forms a hot bubble. The wind is injected from a sphere of radius $R_{\rm inj}=3R_{\sun}$ about the location of the secondary star as it orbits inside the envelope. The hot bubble that we insert can be considered as the cocoon that the jets form (as found by \citealt{MorenoMendezetal2017}).  

We do not change the orbital separation during the simulation. We do not include the gravity of the secondary star, nor do we include the changing gravity due to the deformation of the envelope mass distribution. The gravitational field remains that of the unperturbed AGB star throughout the simulation.

We end the simulation after about 100 days, when the secondary star has completed about four orbits, as some of the assumptions made here, like a constant orbital separation and the omission of the gravity of the secondary star, become too crude. By the end of the simulation the secondary star has accreted about 2 to 10 percents of its initial mass (depending on the efficiency of accretion energy that is carried by the jets). A low-mass secondary star as assumed here becomes fully convective, and is likely to survive as a MS star. 

\subsection {Results}
\label{subsec:results}

We focus on the effects that the hot bubbles have on the envelope and on the mass loss process. For that we will present the flow properties, density, velocity and temperature. When we present the results in the equatorial (orbital) plane, the compact companion moves counterclockwise around the center of the giant star that is located at the center of the grid.
The large black dot is the initial location of the secondary star. 
In all panels the axes are from $-200\, R_{\sun}$ to $200\, R_{\sun}$.
 
\begin{figure} 
\centering
{\includegraphics[width=0.22\textwidth]{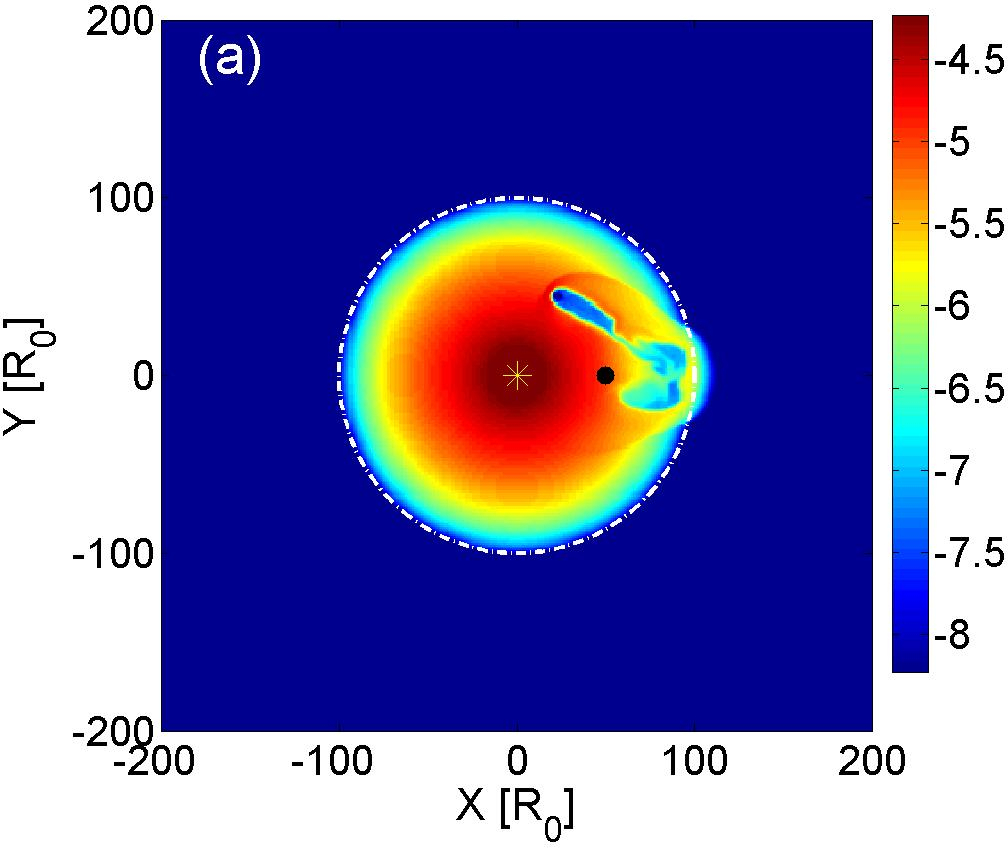}}
\hskip 0.17 cm
{\includegraphics[width=0.22\textwidth]{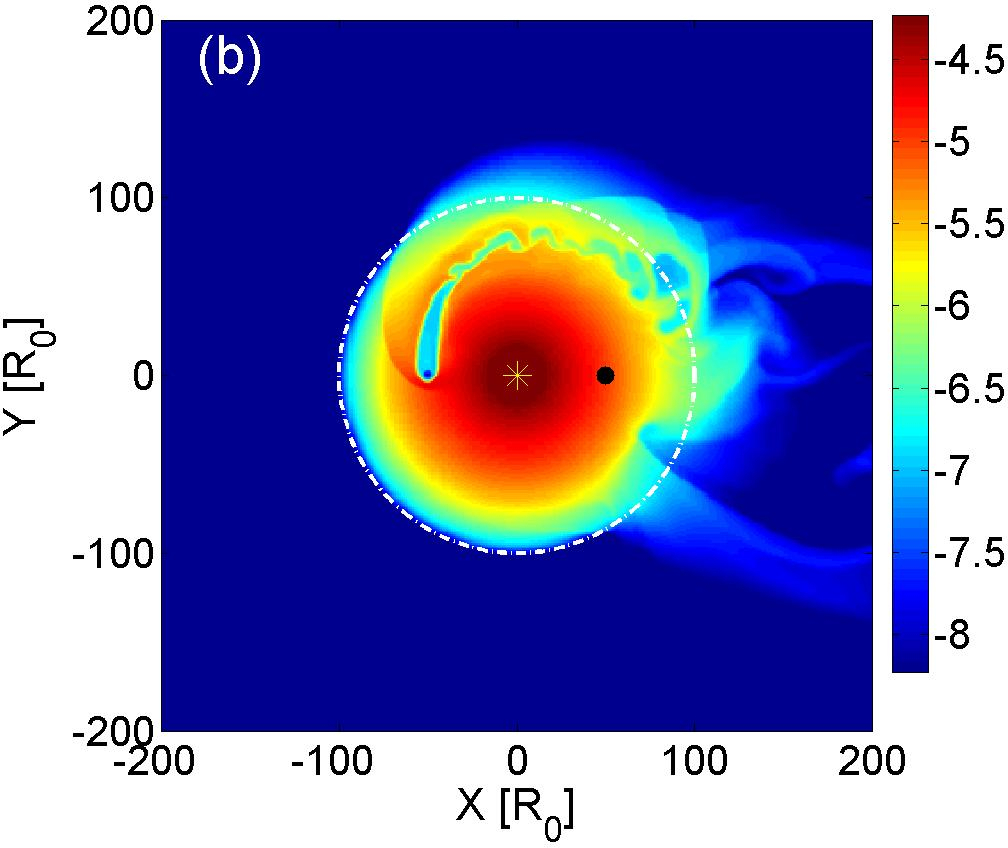}}
{\includegraphics[width=0.22\textwidth]{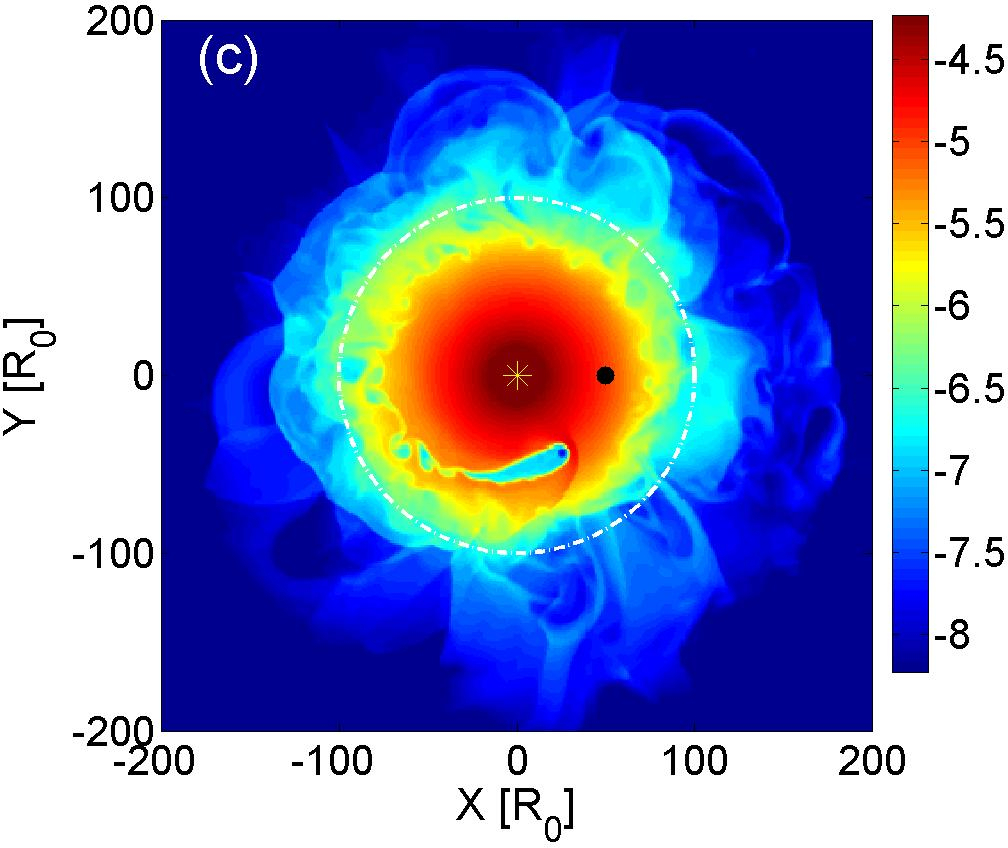}}
\hskip 0.17 cm
{\includegraphics[width=0.22\textwidth]{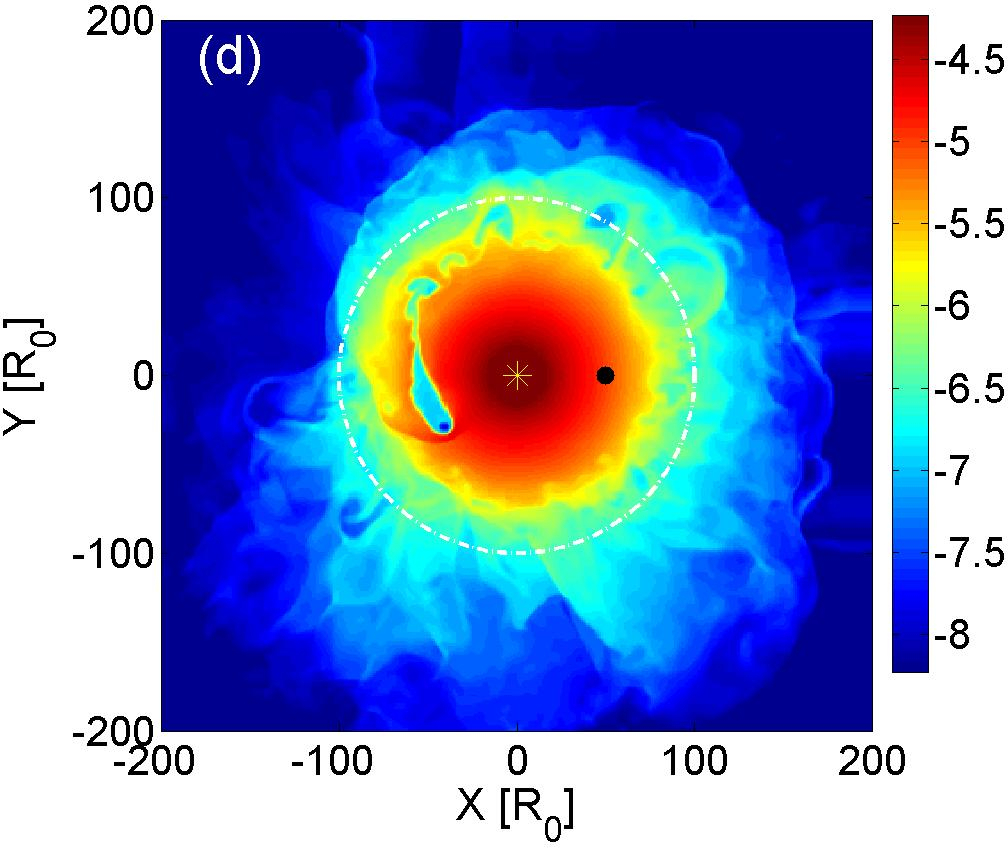}}
{\includegraphics[width=0.22\textwidth]{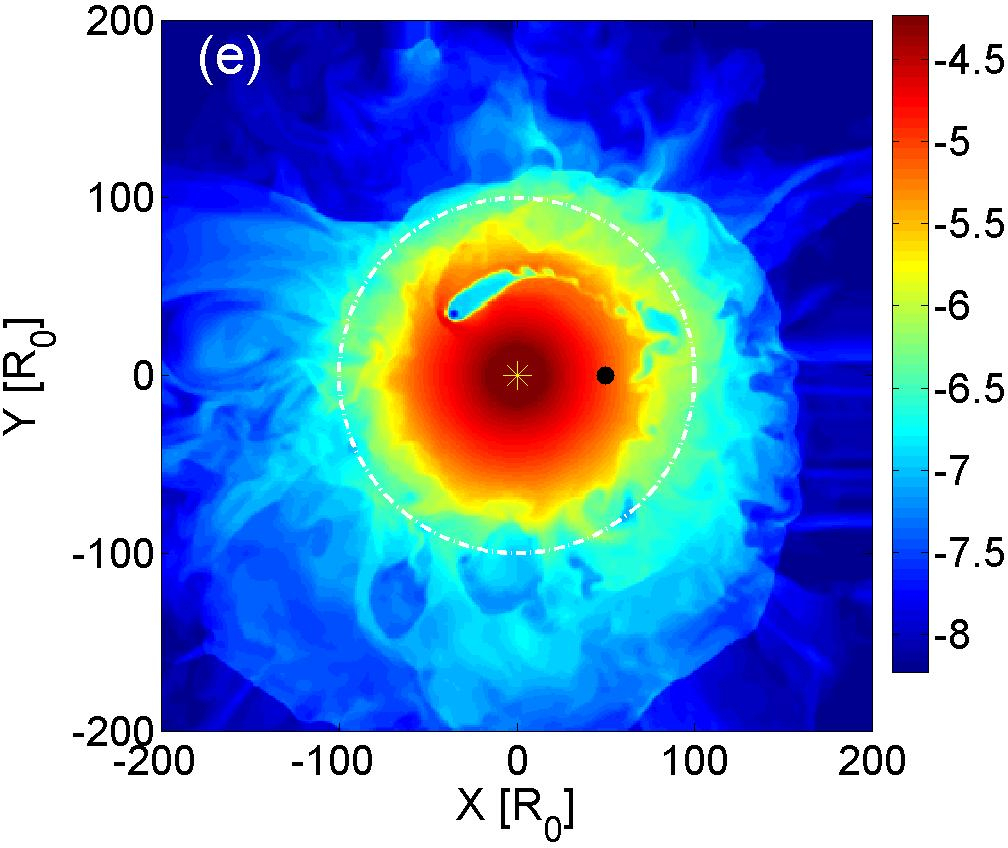}}
\hskip 0.17 cm
{\includegraphics[width=0.22\textwidth]{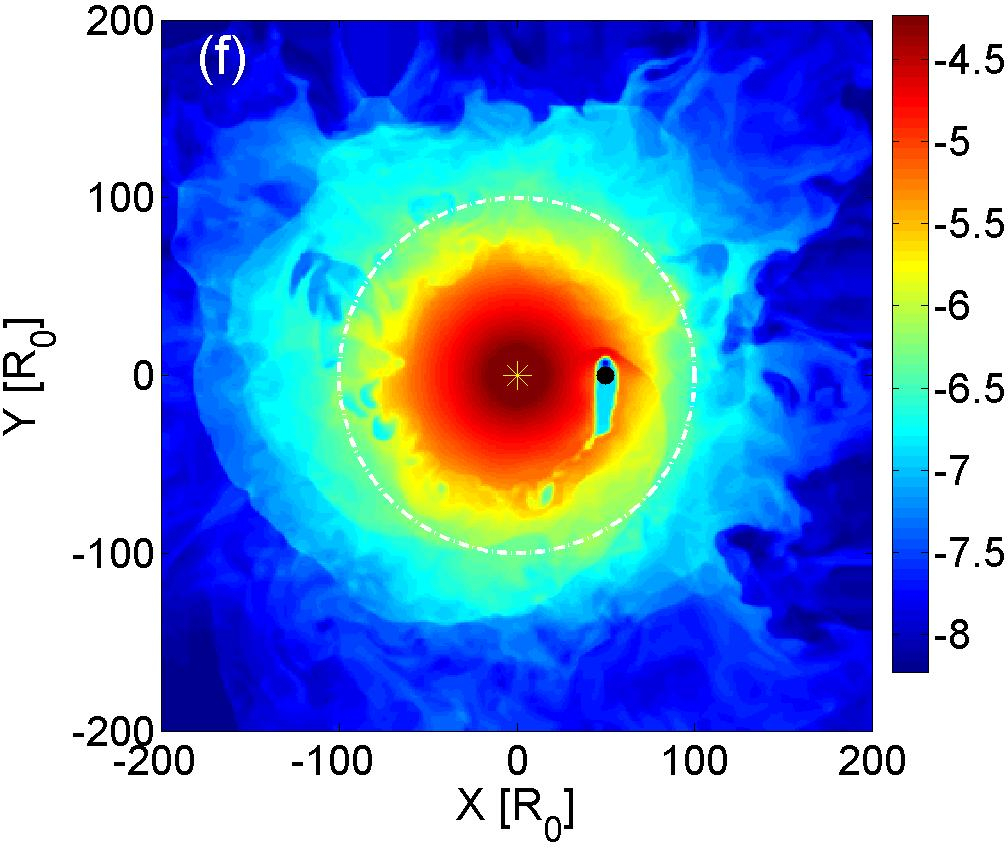}}
\caption{Density maps in the orbital plane $z=0$ at 
six times of (a) $t=4$, 
(b) $12$ (the first half round), 
(c) $46$, (d) $64$, (e) $85$,
and (f) $100 \dday$ (after four rounds), 
from top to bottom and from left to right. The colourmap is in logarithmic scale as indicated by the colourbars and in units of $\g \cm^{-3}$. A yellow asterisk marks the center of the AGB star at the center of the grid. 
A large black dot indicates the initial location of the secondary star.
The dashed-dotted white line marks the initial surface of the AGB star, 
and the small black dot is the location of the secondary,
which is orbiting inside the envelope (counterclockwise). 
The orbital period is $\approx 25 \dday$. 
The last panel shows the density map after four Keplerian orbits.
Units on the axes are in $R_{\sun}$.}
\label{fig:XYslices}
\end{figure}

Fig. \ref{fig:XYslices} presents the density in the equatorial
plane $z=0$ at six times. In panels (a) and (b) we see the formation and initial growth of the bubble. The bubble is the tail behind the secondary star (the small dot at the front of the bubble). In all panels we see that the bubble is unstable and breaks to many small bubbles. At late times we see that after the bubbles break out of the surface they eject mass from the grid.  

\begin{figure} 
\centering
{\includegraphics[width=0.22\textwidth]{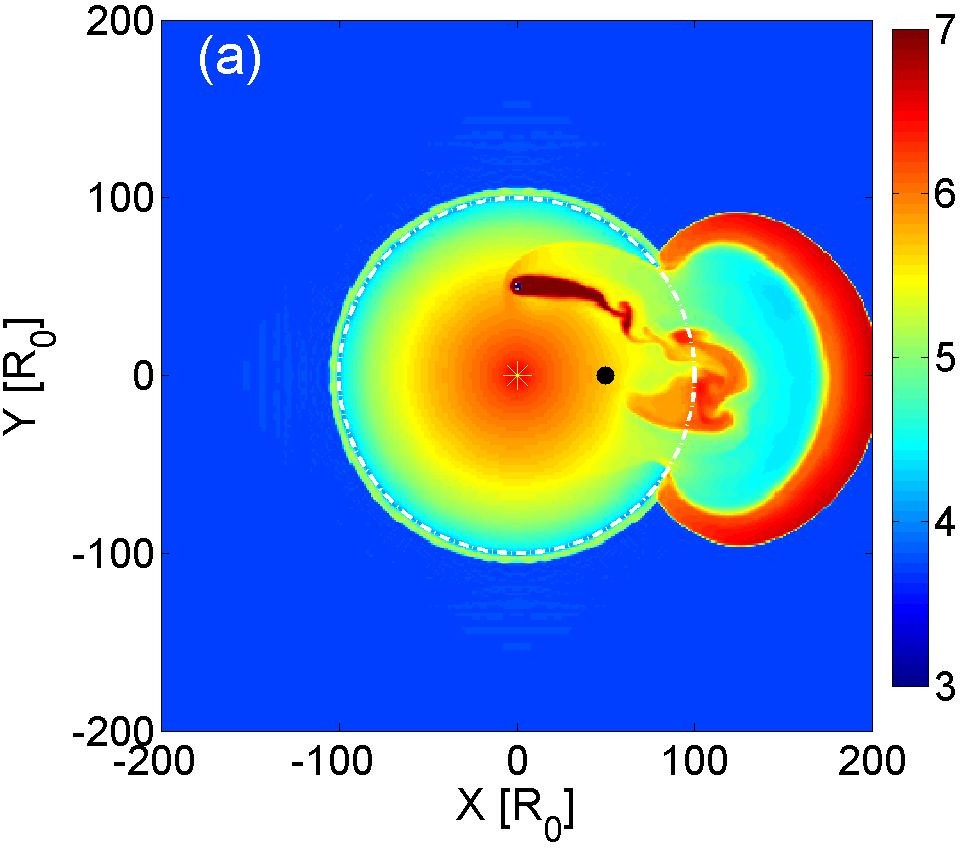}}
\hskip 0.0 cm
{\includegraphics[width=0.22\textwidth]{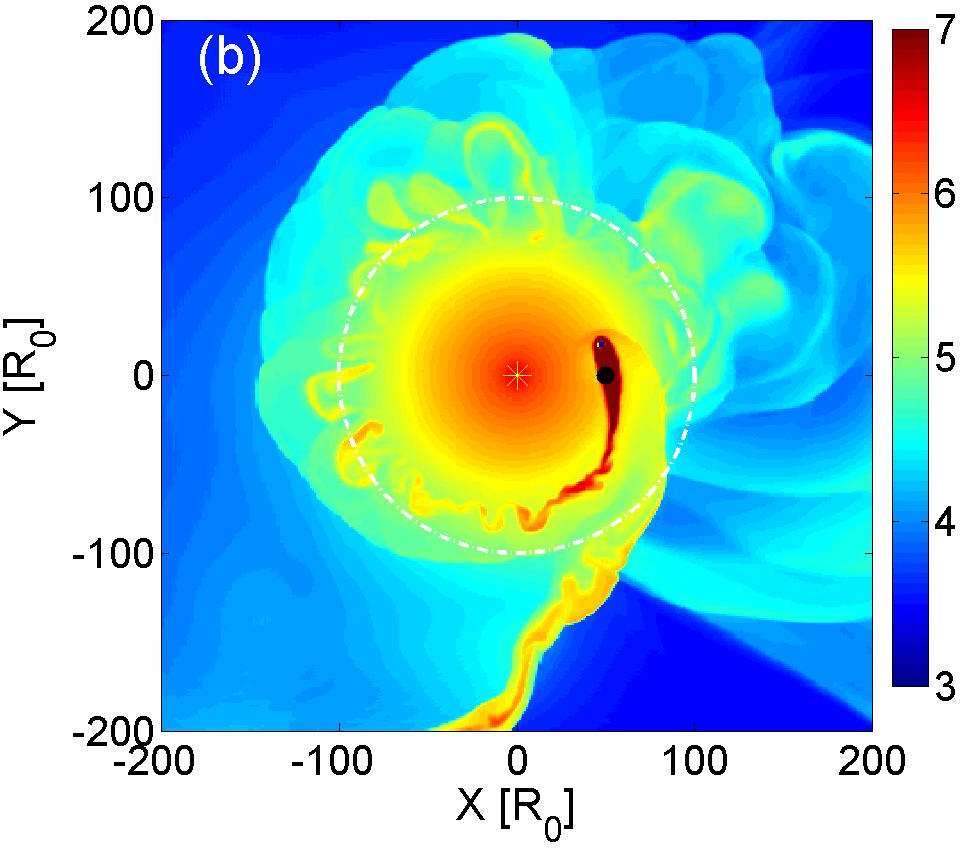}}
{\includegraphics[width=0.22\textwidth]{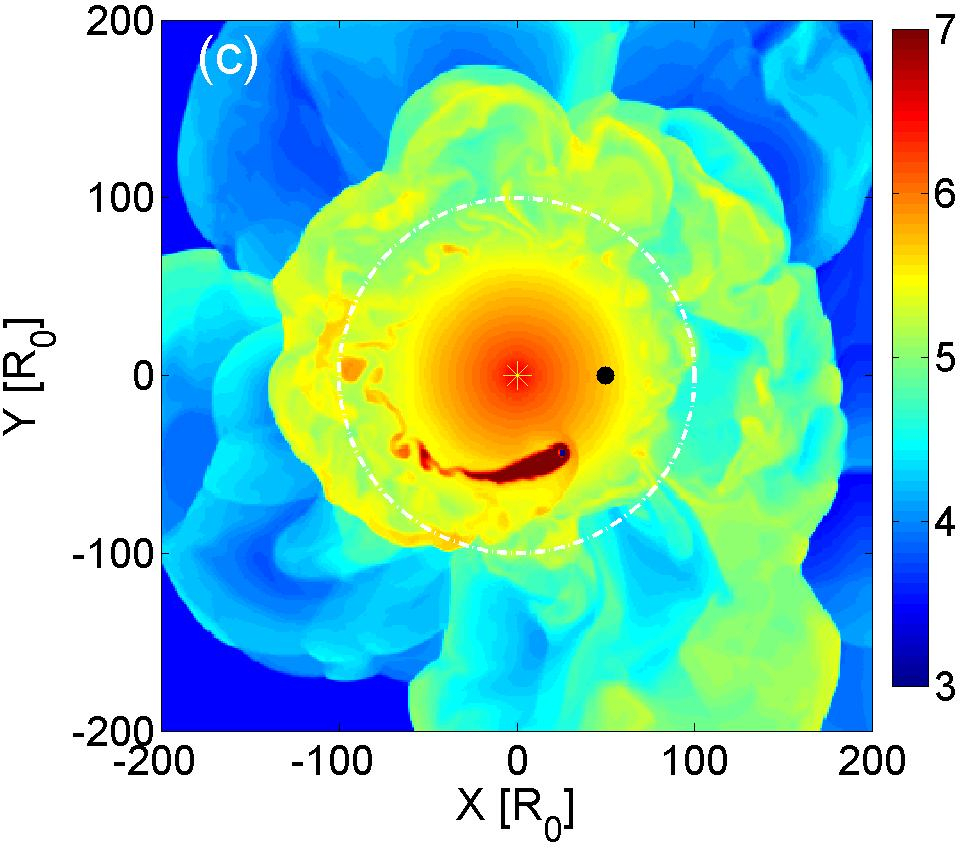}}
\hskip 0.0 cm
{\includegraphics[width=0.22\textwidth]{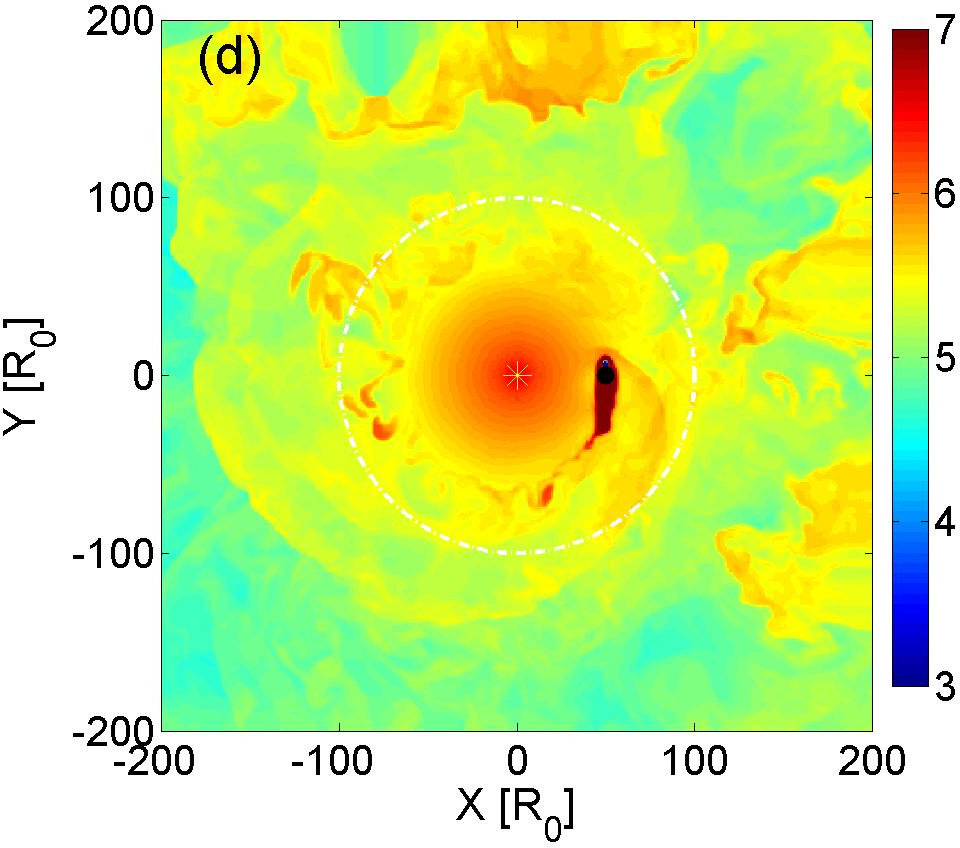}}
\caption{Temperature maps in the orbital plane and at times of 
(a) $t=6$, 
(b) $t=26$, 
(c) $t=46$,
and (d) $t=100 \dday$, colour-coded in a logarithmic scale and in $\K$. }
\label{fig:XYtemperature}
\end{figure}

To further follow the hot bubble and its break-up we present in Fig. \ref{fig:XYtemperature} the temperature in the equatorial plane. 
During the first 30 days there is an effect that results from our initial conditions. We start to inject the outflow from the secondary star when the secondary star is already inside the envelope, rather than follow the entire evolution (due to numerical limitations). This results in a shock that propagates through the very low density region outside the giant, as seen by a red loop in panel (a). This is an artifact that has no influence on the results. The most prominent feature is the hot broken bubble (in red) that trails the secondary star, buoyantly rises through the envelope and ejects mass out of the surface. 
Signatures of the broken bubble are seen as filaments hotter than their surrounding on the left hand side of the star in panel (b).  
The ejection of mass along the orbit results in a spiral pattern, as seen in panels (c) and (d). The immediate temperature of the shocked jets' gas is $\approx 4 \times 10^6 \K$. But the gas is further compressed by the orbital motion, and temperatures rises to $\approx 10^7 \K$.   
 Eventually the bubble cools adiabatically due to expansion, but always stays much hotter than the envelope, unlike convective cells that eventually merge with the envelope.  

\begin{figure} 
\centering
{\includegraphics[width=0.40\textwidth]{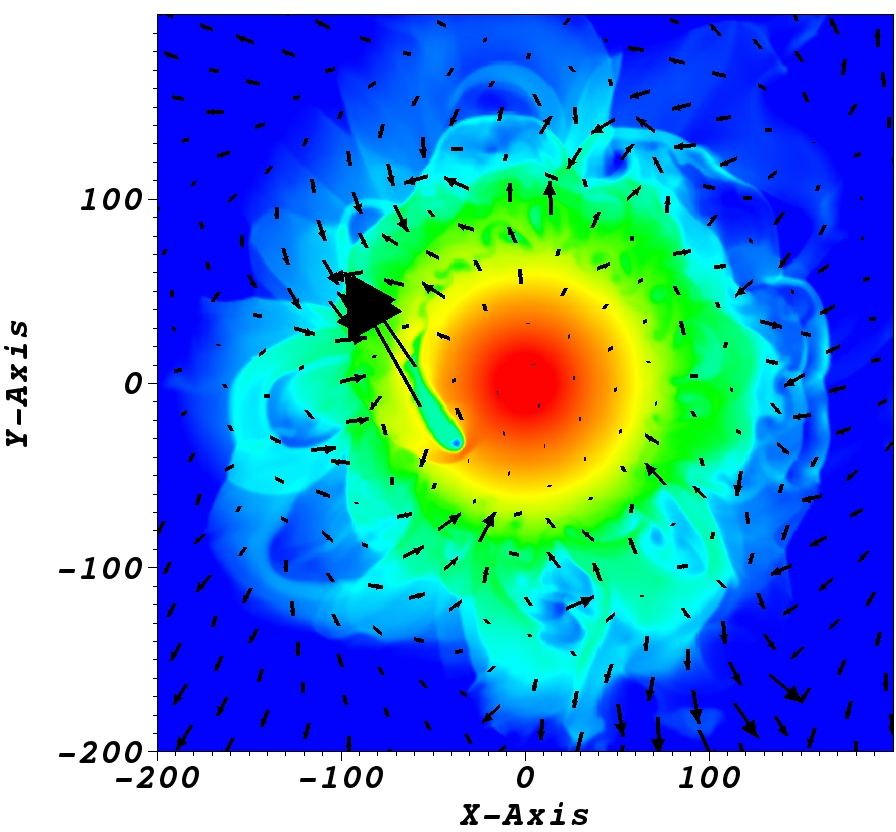}}
{\includegraphics[width=0.40\textwidth]{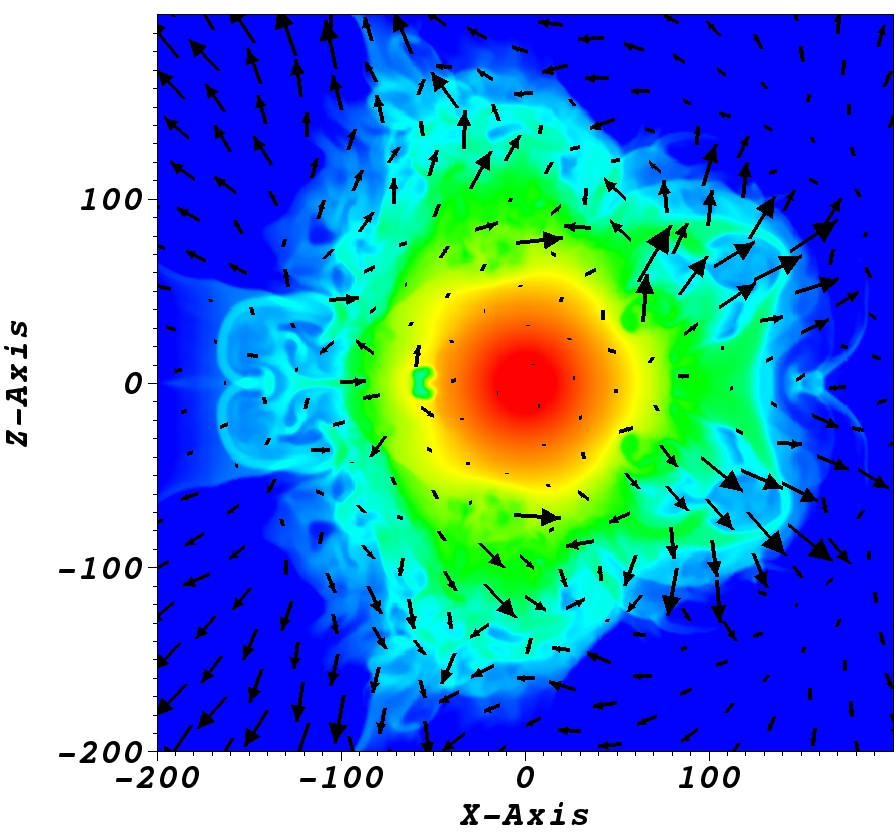}}
\caption{Density and velocity maps at $t=40 \dday$. The top panel is in the orbital plane $z=0$, and the lower one in the meridional plane $y=0$.
The presented density values are from $6\times 10^{-9}$ (blue)
to $6\times 10^{-5}{\rm \, g\,cm^{-3}}$ (red), and in a logarithmic scale. 
The velocity field is annotated by arrows, whose length is proportional to the velocity. The maximum velocities are $555$ and $116\,\km\s^{-1}$, in the upper and lower panels, respectively. Units on the axes are in $R_{\sun}$.  
}
\label{fig:XYvelocity}
\end{figure}

In Fig. \ref{fig:XYvelocity} we present the density and velocity structure at $t=40~$day in the equatorial plane ($z=0$; upper panel) and the meridional plane ($y=0$; lower panel). 
This meridional plane cuts the equatorial plane seen on the upper panel through a horizontal line at the center of the grid which is the center of the giant. The arrows present a complicated flow structure above the stellar surface, including outflows and vortices. The two large arrows represent the injection of the gas that is immediately shocked. Due to the sparse sampling of grid points for arrows, we see only the injection backward, but we inject gas in all directions. As the envelope becomes more disrupted the velocity field becomes more turbulent. 

\begin{figure}
\centering

{\includegraphics[width=0.4\textwidth]{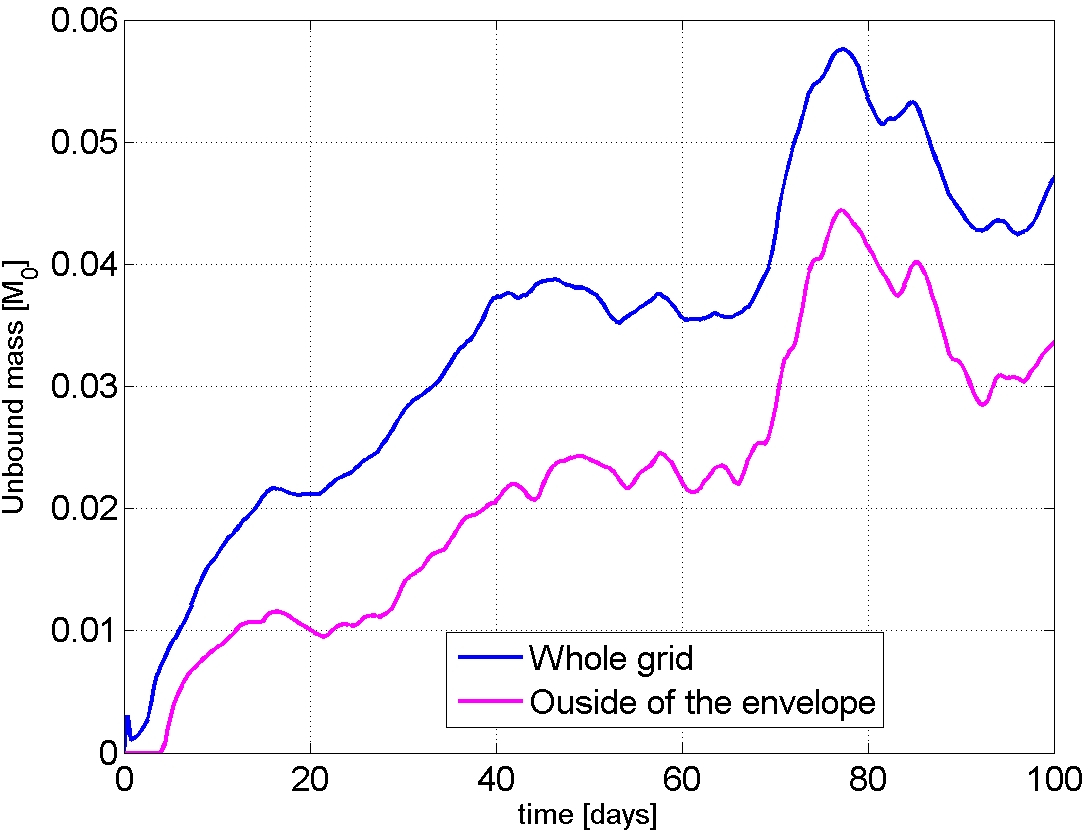}}
\caption{The unbound mass as function of time. The upper blue line is the entire mass in the grid with positive energy, while the lower magenta line is the mass with positive energy outside a radius of $100 R_\odot$ (the initial radius of the star). }
\label{fig:UnboundMass}
\end{figure}

In Fig. \ref{fig:UnboundMass} we present the mass that has sufficient energy to escape the gravitational barrier (total positive energy) as a function of time. The blue line depicts the total unbound mass in the grid, while the magenta line depicts only the unbound mass outside a radius of $100 \,R_{\sun}$. 
At about $t=40~$day a noticeable amount of unbound mass starts to leave the numerical grid, 
as seen here, and in Figs. \ref{fig:XYslices} and \ref{fig:XYvelocity}. Overall, from this graph we deduce an average mass loss rate of $\approx 0.1 M_{\sun} \yr^{-1}$. This is much larger than the mass injection rate into the wind from the secondary star, of $0.0135 M_{\sun} \yr^{-1}$. We also note that the envelope mass leaves the grid in a stochastic manner. This behavior is attributed to the instabilities that break the hot bubble.      

The main result of this section is that jets alone, before the orbital energy is added, can eject mass about ten times the mass in the jets, for the model used here. Again, this is not sufficient to explain the CEE, but we did not consider here the energy released by the in-spiraling binary system. We also note that here we studied an AGB star with a relatively small radius and high mass. 
On the upper AGB the star will be much larger and will lose some envelope in a regular wind. The escape speed from an upper AGB star is lower than in our model, and the jets will be more efficient even in removing envelope gas.

\section{SUMMARY}
\label{sec:summary}

We studied the effect of hot gas in the CEE. First we examined the suggestion that the recombination energy of hydrogen and helium removes the common envelope. 
In section \ref{sec:radiative} we showed that the optical depth from the hydrogen recombination zone to the photosphere is low, such that a large fraction of the recombining photons diffuse out in a time shorter than the envelope ejection time \citep{Harpaz1998}. 
In section \ref{sec:convective} we used simple arguments to claim that the convection can carry a large fraction of the recombination energy of helium to the surface, where it is radiated away (section \ref{sec:radiative}). 
In reality, the envelope expands and its density drops as a result of energy deposition by the spiraling-in process and by jets if launched by the companion. This on the one hand reduces the photon diffusion time outwards, but on the other hand the lower density makes convection less efficient. The overall effects of jets activity and the spiraling-in process on the fraction of the recombination energy that does not escape in radiation, should be studied by numerical simulations that include all processes, including radiative transfer.

Overall, we concluded in sections \ref{sec:radiative} and \ref{sec:convective} that the recombination energy does not contribute much to the removal of the envelope. The recombination energy is mainly radiated away, adding to the luminosity of a transient event that might be observed (i.e., an ILOT).  

In light of our finding and in light of other difficulties in removing the envelope encountered in simulations of the CEE (section \ref{sec:intro}), there is a need to consider other extra energy source(s). We here studied the possible effects of jets launched by the secondary star (section \ref{subsec:preface} and review by \citealt{Soker2016Rev}). 
Due to numerical limitations we injected an isotropic wind instead of jets (section \ref{subsec:scheme}). We described our results in section \ref{subsec:results}. 

We summarize the main results of our 3D simulation as follows. 
Unlike convective cells that rise a limited distance and merge with the envelope, the hot bubble that is formed by the secondary star rises all the way to the surface, heavily disturbs the envelope, and ejects some envelope mass. 
The bubble itself breaks-up to many small bubbles that lead to a complicated behavior of the flow and mass ejection. For our specific model and omission of the orbital gravitational energy, the jets eject about ten times their own mass. 
The energy that is released by the in-spiraling binary system causes a huge envelope expansion (see references to relevant numerical simulations in section \ref{sec:intro}). 
Had we included the gravitational energy of the spiraling-in binary system, the ejected mass would be much larger. This is a topic of a future study. 

More generally, our results point out that it is crucial to include energy transport by convection during the common envelope process. 

\section*{Acknowledgments}
We thank an anonymous referee for very helpful comments.
We thank Michael Refaelovich for his assistance. 
This research was supported by the Prof. A. Pazy Research Foundation.





\label{lastpage}
\end{document}